\documentclass{aa}
\usepackage[varg]{txfonts}
\usepackage[dvipsnames]{xcolor}
\usepackage{graphicx}
\usepackage{siunitx}

\bibpunct{(}{)}{;}{a}{}{,} 

\begin{document}



\title{Kinetic-Scale Physics of a Multi-Species Solar Wind}
\subtitle{Interaction with a Comet}
\author{Anja Moeslinger\inst{1, 2}
	\and Herbert Gunell\inst{2}
	\and Gabriella Stenberg Wieser\inst{1}
	\and Hans Nilsson\inst{1}
	\and Shahab Fatemi\inst{2}}
\institute{Swedish Institute of Space Physics, Kiruna, Sweden
	\and Department of Physics, Ume{\aa} University, Ume{\aa}, Sweden}
\date{Received now / Accepted later}
\abstract {The solar wind affects the plasma environment around all solar system bodies. A strong solar wind dynamic pressure pushes plasma boundaries closer to these objects. For small objects kinetic effects on scales smaller than an ion gyroradius play an important role, and species with various mass-per-charge may act differently. In this case the solar wind composition can be important. } 
{
Protons are the dominant ion species in the solar wind; however,  
sometimes the density of alpha particles increases significantly.
We analyse the effect of different solar wind alpha-to-proton ratios on the plasma boundaries of the induced cometary magnetosphere. In addition, we investigate the energy transfer between the solar wind ions, the cometary ions, and the electromagnetic fields.} 
{Using the hybrid model Amitis, we simulate two different alpha-to-proton ratios and analyse the resulting plasma structures. 
We calculate the power density ($\mathbf{E\cdot J}$) of all three ion species (solar wind protons and alphas, and cometary ions) to identify load and generator regions. 
The integrated 1D power density shows the evolution of the power density from the upstream solar wind to downstream of the nucleus. 
} 
{A higher alpha-to-proton ratio leads to a larger comet magnetosphere but weaker magnetic field pile-up. The protons transfer energy to the fields and the cometary ions in the entire upstream region and the pile-up layer. Upstream of the nucleus, alphas are inefficient in transferring energy and can act as a load, especially for low alpha-to-proton ratios. The transfer of energy from alphas to cometary ions happens further downstream due to their larger inertia.} 
{For a multi-species solar wind the mass loading and energy transfer upstream of the pile-up layer will be most efficient for the species with the lowest inertia, typically protons, since different ion gyroradii give different flow patterns for the individual species.
}
\maketitle




\section{Introduction}
Long and structured comet tails are stunning displays in the sky, sometimes even visible by the naked eye. Even before the first measurements of the solar wind (SW) was made, observations of comets suggested the presence of a stream of particles originating from the Sun \citep{biermann1951za} as well as the work of electromagnetic forces \citep{alfven1957tellus}. The partially ionised cometary atmosphere presents an obstacle to the solar wind plasma flow, and an induced magnetosphere forms around the nucleus. As the outgassing rates vary enormously, both between different comets and along the comets' elliptical orbits around the Sun, different regimes of SW interaction can be identified \citep{goetz2022ssr}. At a high-activity comet a fluid approach captures most of the large scale features of the magnetosphere \citep{cravens2004asr}, while at a low-activity comet kinetic effects play major roles \citep{nilsson2018aa}. 

The Rosetta mission \citep{glassmeier2007ssr} to comet 67P/Churyumov-Gerasimenko put focus on the latter interaction domain, where the typical scales of the magnetosphere are smaller than an ion gyroradius. In the low-activity case, an asymmetric magnetosphere is created \citep{behar2016aa, behar2018baa}, with clear boundaries appearing only in one hemisphere. A bow shock is likely not present but the first signs of shock formation are occasionally observed \citep{gunell2018aa}. The solar wind ions are more deflected than decelerated when they encounter the cometary plasma. In this regime, both solar wind protons and cometary ions exhibit highly anisotropic distributions, at times resembling partial rings \citep{moeslinger_vdfs_2024, moeslinger2023ajgr, moeslinger2023bjgr, williamson2022aa}. 

The SW-cometary plasma interaction is sensitive not only to the properties of the comet itself but also to the state of the solar wind. For example, a change in the interplanetary magnetic field due to a crossing of the heliospheric current sheet can lead to a disconnection of the cometary tail \citep{jia2007jgr}. The impact of a Coronal Mass Ejection (CME) may have the same result \citep{vourlidas2007apj}, and \citet{edberg2016mnras} reported dramatic changes also in the inner coma of comet 67P during a CME event. The compression of the magnetosphere and increased ionisation lead to higher densities, stronger magnetic fields, and the presence of magnetic flux ropes.

Alpha particles are the second most abundant SW species after protons, with typical relative number densities $n_\alpha/n_H$ between $3-\SI{5}{\percent}$ in the steady-state SW. There is a strong dependence of $n_\alpha/n_H$ with solar activity \citep{kasper_2007, song_2022}. The lowest $n_\alpha/n_H~(< \SI{1}{\percent})$ occurs during solar minimum in the fast solar wind, while $n_\alpha/n_H > \SI{20}{\percent}$ is possible during coronal mass ejections (CMEs) at solar maximum.  
Additionally, \citet{yogesh_2023} have shown that $n_\alpha/n_H$ also increases rapidly at the surface of stream interaction regions (SIRs).
Close to the Sun alphas typically have higher bulk velocities than protons (up to \SI{150}{\km\per\s} higher at \SI{0.3}{AU}), but this difference decreases with increasing heliocentric distance (e.\,g.~ \SI{30}{\km\per\s} at \SI{1}{AU}, \SI{20}{\km\per\s} at $2.5 - \SI{3}{AU}$) \citep{reisenfeld_helium_2001, durovcova_2017}. This differential flow can cause instabilities like electron cyclotron waves \citep{zhao_2019}.

An instructive approach to studying the interaction between the solar wind and the cometary coma is using Poynting's theorem to focus on the energy transfer between the different particle populations and the electromagnetic fields.
Poynting's theorem expresses the conservation of electromagnetic energy \citep{olsson1987book} and can be written as
\begin{equation}
	-\frac{\partial}{\partial t}\left(\frac{B^2}{2 \mu_0}+\frac{\varepsilon_0{E^2}}{2}\right) = (\mathbf{\nabla \cdot S + E \cdot J}),
	\label{eq:poynting}
\end{equation}
where $B = |\mathbf{B}|$ and $E=|\mathbf{E}|$ are the magnitudes of the magnetic and electric fields, respectively, $\mu_0$ is the permeability of vacuum, and $\varepsilon_0$ is the vacuum dielectric constant. Hence, the left hand side is the change over time of the energy density stored in the electromagnetic fields. Often ${B^2}/{2 \mu_0}\gg{\varepsilon_0{E^2}}/{2}$, and ${\varepsilon_0{E^2}}/{2}$ can be neglected. To conserve energy the change in electromagnetic energy density is compensated by a flow of field energy to or from the region of interest, given by $\mathbf{\nabla \cdot S}$, where $\mathbf{S = E \times B}/\mu_0$ is the Poynting flux (Poynting vector). If charged particles are present in the system, energy can also be transferred to or from them through work done by the Lorenz force. The term $\mathbf{E \cdot J}$, where $\mathbf{J}$ is the current density, represents this energy transfer. A power density $\mathbf{E \cdot J}>0$ means that electromagnetic energy is given to the charge carriers, that is, the power density is a ``load''. A negative power density ($\mathbf{E \cdot J}<0$) is a ``generator'', where energy moves from the charged particles to the electromagnetic fields.

Studies of many different objects in the solar system make use of Poynting's theorem for the interpretation of their data in terms of energy conversion. \citet{saunders1986jgr} explain measurements made in Venus' magnetotail by investigating the sign of $\mathbf{E \cdot J}$ in a simple model based on the magnetic field observations. Multi-spacecraft constellations made direct observations of the power density easier as the current density can be computed as $\mathbf{J = \nabla \times B}/\mathrm{\mu_0}$. A number of papers make use of this to identify generator and load regions related to the auroral current circuit at Earth and energy conversion in the plasma sheet \citep{hamrin2006ag, marghito2006ag, hamrin2012jgr}. In an investigation of the solar wind input to the Earth's magnetosphere \citet{lockwood2019jgr} show that including the solar wind Poynting flux to the total power input improves the correlation with the \textit{am} geomagnetic index on time scales of a day and shorter.

The data needed to compute the different terms in Poynting's theorem is often straightforward to extract in numerical models. The power transfer to different ion species in the Martian induced magnetosphere is investigated by \citet{wang2024mnras} using a hybrid plasma model. They show that both the bow shock and the induced magnetosphere boundary are generator regions, while the planetary ions act as a load and gain energy from the electromagnetic fields. \citet{lindkvist_energy_2018} use a hybrid model and Poynting's theorem to study the conversion of energy inside the magnetosphere of comet 67P at a heliocentric distance of \SI{1.5}{AU} for two extreme conditions of solar EUV radiation. They find an asymmetric magnetosphere, where the region of the initial pile-up of the interplanetary magnetic field is a generator and the cometary water ions act as a load, similar to what is seen in the Martian model. In \citet{lindkvist_energy_2018} a single ion solar wind is used, consisting only of protons and electrons. In this paper we include a second solar wind ion species, alpha particles, in a hybrid model and investigate the effects a multi-ion SW has on the energy conversion at a comet by simulating two extreme values of the alpha to proton ratio under otherwise constant conditions.

\section{Methods}

Hybrid models are frequently used to model plasma processes where kinetic effects of ions are important. The main tool used in this study is the hybrid model code Amitis\footnote{https://www.amitiscode.com/}.
Amitis \citep{fatemi_amitis_2017} is a GPU-based implementation of the hybrid model and has been successfully used to model numerous objects, such as the plasma environment of  Mercury \citep{fatemi_mercury_2018, aizawa_cross-comparison_2021}, comets \citep{gunell_radial_2024}, Ganymede inside Jupiter's magnetosphere \citep{fatemi_ganymede_2022}, and recently presented the first 3D hybrid-kinetic simulations of the solar wind interaction with the entire magnetosphere of Earth at physical scales \citep{fatemi_jets_2024}.

In hybrid simulations, ions are modelled as so-called macro particles, where one such macro particle represents a large number of ``physical'' particles, scaled by the so-called macro particle weight. Position and velocity of each macro particle are computed at every time step using the Lorentz equation of motion \citep{fatemi_amitis_2017}. Multiple species of ions, in our case SW protons, SW alphas, and cometary ions (H$_2$O$^+$, \SI{18}{amu}), are simulated. Electrons are implemented as a massless, charge-neutralising fluid, enforcing quasi-neutrality in the entire simulation domain.
The simulation domain consists of a cartesian grid of size $\num{7000}\times\num{8800}\times\num{22000}\,\si{km}$ $(x\times y\times z)$ and is split up into cubic cells with a grid resolution of $\delta x = \delta y = \delta z = \SI{40}{km}$. The electric and magnetic fields ($\mathbf{E}$ and $\mathbf{B}$) are calculated for each grid cell.
The $x-$axis of the simulation coordinate system points towards the sun, and the upstream SW velocity $\mathbf{v}_\mathrm{SW}$ is along the $-x$-axis. The interplanetary magnetic field $\mathbf{B}_\mathrm{SW}$ is perpendicular to the SW velocity and is oriented along the $y-$axis. The convective electric field of the SW $\left( \mathbf{E}_\mathrm{SW} = -\mathbf{v}_\mathrm{SW} \times \mathbf{B}_\mathrm{SW} \right)$ is therefore along the $z-$axis and splits the simulation domain into a $+E$ ($z\geq 0$) and a $-E$ ($z<0$) hemisphere.

The ion number and charge density ($n_i$ and $\rho_i$) and ion current density $\mathbf{J}_i$ can be calculated from the distribution of macro particles at each time step. The electric current density $\mathbf{J}$ is calculated by Ampere's law ignoring the displacement current (Darwin limit):
\begin{align}
	\mathbf{J} = \frac{1}{\mu_0} \nabla \times \mathbf{B} = \mathbf{J}_i + \mathbf{J}_e.
	\label{eq::J_hybrid}
\end{align}
This also gives the electron current $\mathbf{J}_e$ and the electron velocity $\mathbf{v}_e$. The electric field is given by Ohm's law:
\begin{align}
	\mathbf{E} = -\mathbf{v}_i \times \mathbf{B} + \frac{\mathbf{J}}{\rho_i} \times \mathbf{B} - \frac{\nabla p_e}{\rho_i} + \eta \mathbf{J} = -\mathbf{v}_e \times \mathbf{B} - \frac{\nabla p_e}{\rho_i} + \eta \mathbf{J}
	\label{eq::E_hybrid}
\end{align}
with the ion velocity $\mathbf{v}_i = \mathbf{J}_i/\rho_i$, the electron pressure $p_e \propto T_e n_i^\gamma, \gamma=5/3$, $T_e = \SI{1.3e5}{\K}$ (electron temperature), and the plasma resistivity $\eta = \SI{5.0e3}{\ohm\m}$. The plasma resistivity helps to dampen numerical oscillations in areas with strong field gradients. The ions are moved using the non-dissipative electric field $\mathbf{E}_\mathrm{ND}$ which does not include the resistive term:
\begin{align}
	\mathbf{E}_\mathrm{ND} = \mathbf{E} - \eta\mathbf{J} = -\mathbf{v}_i \times \mathbf{B} + \frac{\mathbf{J}}{\rho_i} \times \mathbf{B} - \frac{\nabla p_e}{\rho_i}
	\label{eq::E_ND_hybrid}.
\end{align}
Faraday's law is used to propagate the magnetic field: 
\begin{align}
	\frac{\partial\mathbf{B}}{\partial t} = -\nabla\times\mathbf{E}.
	\label{eq::faraday}
\end{align}
If there are not enough macro particles in a grid cell ($\rho_i \rightarrow 0$), Equation \ref{eq::faraday} is approximated by the magnetic diffusion equation:
\begin{align}
	\frac{\partial\mathbf{B}}{\partial t} = \frac{\eta_\mathrm{vac}}{\mu_0}\, \nabla^2\mathbf{B}
	\label{eq::B_vac}
\end{align}
where we use the vacuum resistivity $\eta_\mathrm{vac} = \SI{2.5e6}{\ohm\m}$. This also defines the minimum time step for the simulation: $\Delta t \leq \mu_0 (\Delta x)^2/\left(2\eta_\mathrm{vac}\right) = \SI{0.4}{ms}$. In our simulations including the vacuum resistivity is relevant during the startup phase of the simulation. Once the simulation reaches a quasi-steady-state no vacuum regions are present.
More information about hybrid models can be found e.\,g.~in \citet{ledvina_modeling_2008}. Details about Amitis can be found in \citet{fatemi_amitis_2017}.

We simulate two different SW compositions, and the corresponding simulation runs are subsequently referred to as \emph{low alpha} and \emph{high alpha} case. The major difference between those two runs are the proton and alpha particle densities in the upstream SW. Density values and ratios of the two species are listed in Table \ref{tab::simulation_parameters}. The alpha-to-proton ratios were chosen to represent extremes of the SW composition. For a comparable comet-SW interaction the total upstream SW dynamic pressure is conserved between the two simulations. Since both SW species have the same upstream velocity, this also corresponds to a conservation of SW mass density, and SW momentum between the two simulations. However, the SW plasma density is lower in the high alpha case.

Most plasma parameters in our simulations roughly correspond to the plasma environment of comet 67P at a heliocentric distance of $2.5 - \SI{3}{AU}$.
At every time step SW particles are injected into the simulation domain at the upstream boundary at $x=+\SI{5000}{km}$ with the defined SW parameters following a Maxwell velocity distribution (see Table \ref{tab::simulation_parameters}).
In both simulations, SW protons and SW alphas have 40 ppc each. The total number of particles -- including the solar wind and cometary -- in the simulation is $\approx 1.8$ billion, which are individually tracked at every time step.
The cometary neutral density is based on a spherically symmetric outgassing profile with outgassing rate $Q$ \citep{haser_1957} and constant neutral expansion velocity $u_n = \SI{700}{\m\per\s}$ \citep{gulkis_subsurface_2015, lee_spatial_2015}. Neutrals are ionised by photoionisation with the photoionisation rate $\nu^{h\nu, \mathrm{ioni}}$, and the ion production profile as a function of radial distance $r$ is:
\begin{align}
	P(r) = \frac{Q \nu^{h\nu, \mathrm{ioni}}}{4\pi u_n r^2}
	\label{eq::ion_production}.
\end{align}
with $Q \nu^{h\nu, \mathrm{ioni}} = \SI{1.08e26}{\per\s\squared}$ \citep{hansen_evolution_2016, heritier_plasma_2018}.
The cometary ions are produced at every time step based on equation \ref{eq::ion_production}. 

\begin{table}[hbt]
	\caption{Upstream SW parameters for both simulation runs.}
  	\begin{tabular}{l l l}
    	Parameter & low alpha & high alpha \\
    	\hline
    	magnetic field strength $B_\mathrm{SW}$ & \SI{3}{nT} & \SI{3}{nT} \\
    	solar wind speed $v_\mathrm{SW}$ & \SI{430}{\km\per\s} & \SI{430}{\km\per\s} \\
    	proton density $n_H$ & \SI{1.156}{\per\cm\cubed} & \SI{0.667}{\per\cm\cubed} \\
    	alpha density $n_\alpha$ & \SI{0.0115}{\per\cm\cubed} & \SI{0.133}{\per\cm\cubed} \\
    	$\alpha/H$ number ratio & \SI{1}{\percent} & \SI{20}{\percent} \\
    	$\alpha/H$ mass density ratio & \SI{4}{\percent} & \SI{80}{\percent} \\
    	$\alpha/H$ charge density ratio & \SI{2}{\percent} & \SI{40}{\percent} \\
  \end{tabular}
  \label{tab::simulation_parameters}
\end{table}

The total electromagnetic power density $\mathbf{E\cdot J}$ can be directly calculated from the model output variables for each grid point. To calculate the power density of the individual ion species we first calculate the non-dissipative electric field $\mathbf{E}_\mathrm{ND}$ used to move the macro particles in the simulation (see equation \ref{eq::E_ND_hybrid}). The power density of species $n$ is then calculated as $\mathbf{E}_\mathrm{ND}\cdot\mathbf{J}_{i, n}$, where $\mathbf{J}_{i, n}$ is the current density of ion species $n$. The total $\mathbf{E\cdot J}$ can be split up in three components:
\begin{align}
	\mathbf{E\cdot J} = \left( \mathbf{E}_\mathrm{ND} + \eta \mathbf{J} \right) \cdot \left( \mathbf{J}_i + \mathbf{J}_e \right) = \mathbf{E}_\mathrm{ND} \cdot \mathbf{J}_i + \mathbf{E}_\mathrm{ND} \cdot \mathbf{J}_e + \eta J^2.
	\label{eq::EdotJ}
\end{align}
The first term on the right-hand side is the summed power density of all ion species and is non-dissipative in the simulation. The second and third term make up the power density of the electron fluid. The second term is the non-dissipative part of the electron power density, while the third term is dissipative and describes processes like heating of the electron fluid. We do not analyse the electron power density in this paper, however, we want to emphasise that the presented power densities are not expected to sum up to zero.

\section{Results}
\subsection{Spatial structure}
\begin{figure}
	\resizebox{\hsize}{!}{\includegraphics{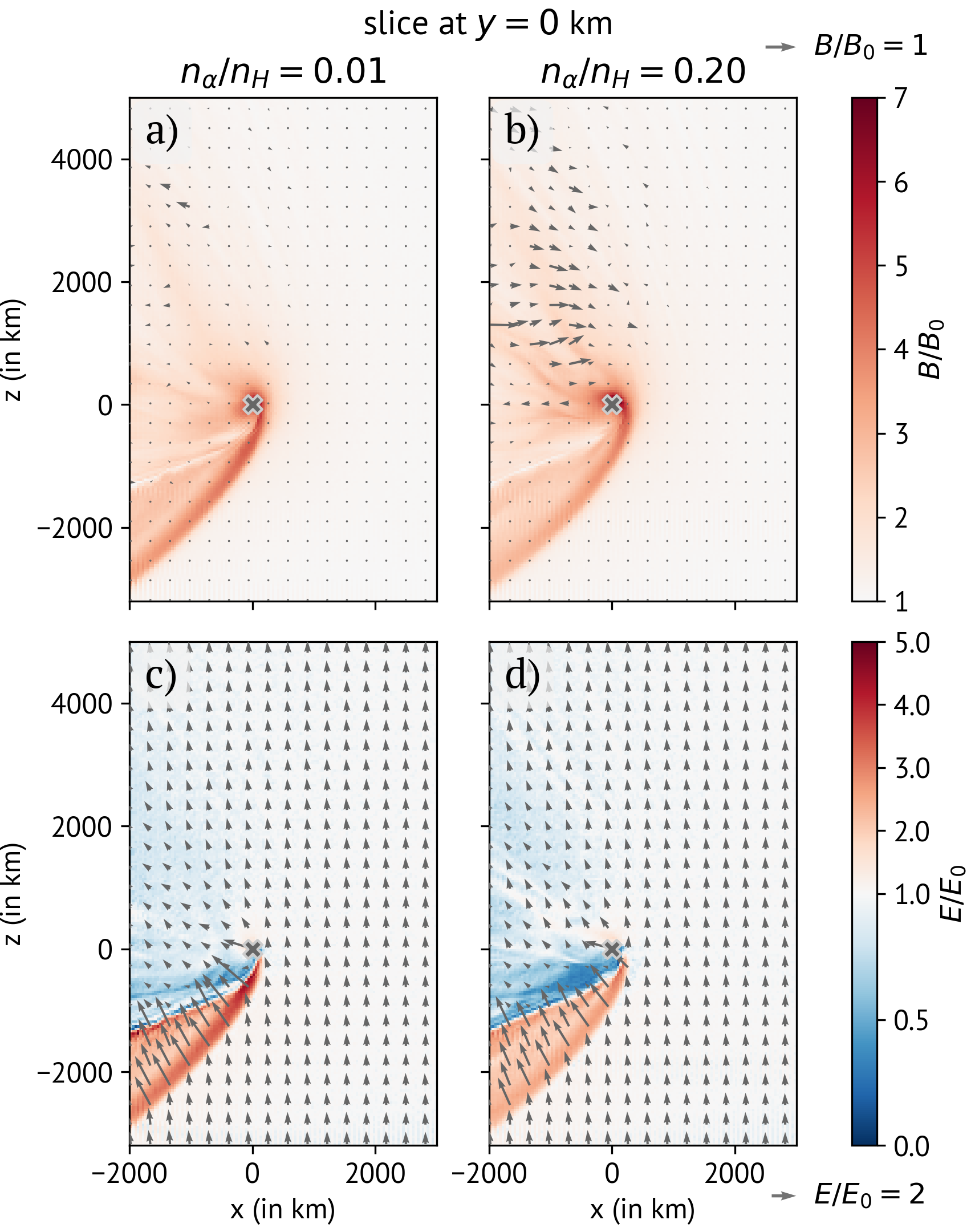}}
	\caption{Comparison of the electric and magnetic fields between the low (left column, panels a and c) and high (right column, panels b and d) alpha simulations. All panels show a slice of the $x-z$ plane, at $y=0$. Top row (panels a and b): magnetic field, normalised to the upstream value $(B_0 = \SI{3}{nT})$. Bottom row (panels c and d): normalised electric field $(E_0 = \SI{1.29}{mV \per m})$. The background colour shows the magnitude of the electric and magnetic field. The arrows show the strength and direction of the fields in this plane, that is, $x-$ and $z-$component only. The length scales for the arrows are specified above (magnetic field) and below (electric field) the respective colour bars on the right side. The comet is located at $(0, 0)$ and marked with a grey `x'. }
	\label{fig::E+B_y0}
\end{figure}

\begin{figure}
	\resizebox{\hsize}{!}{\includegraphics{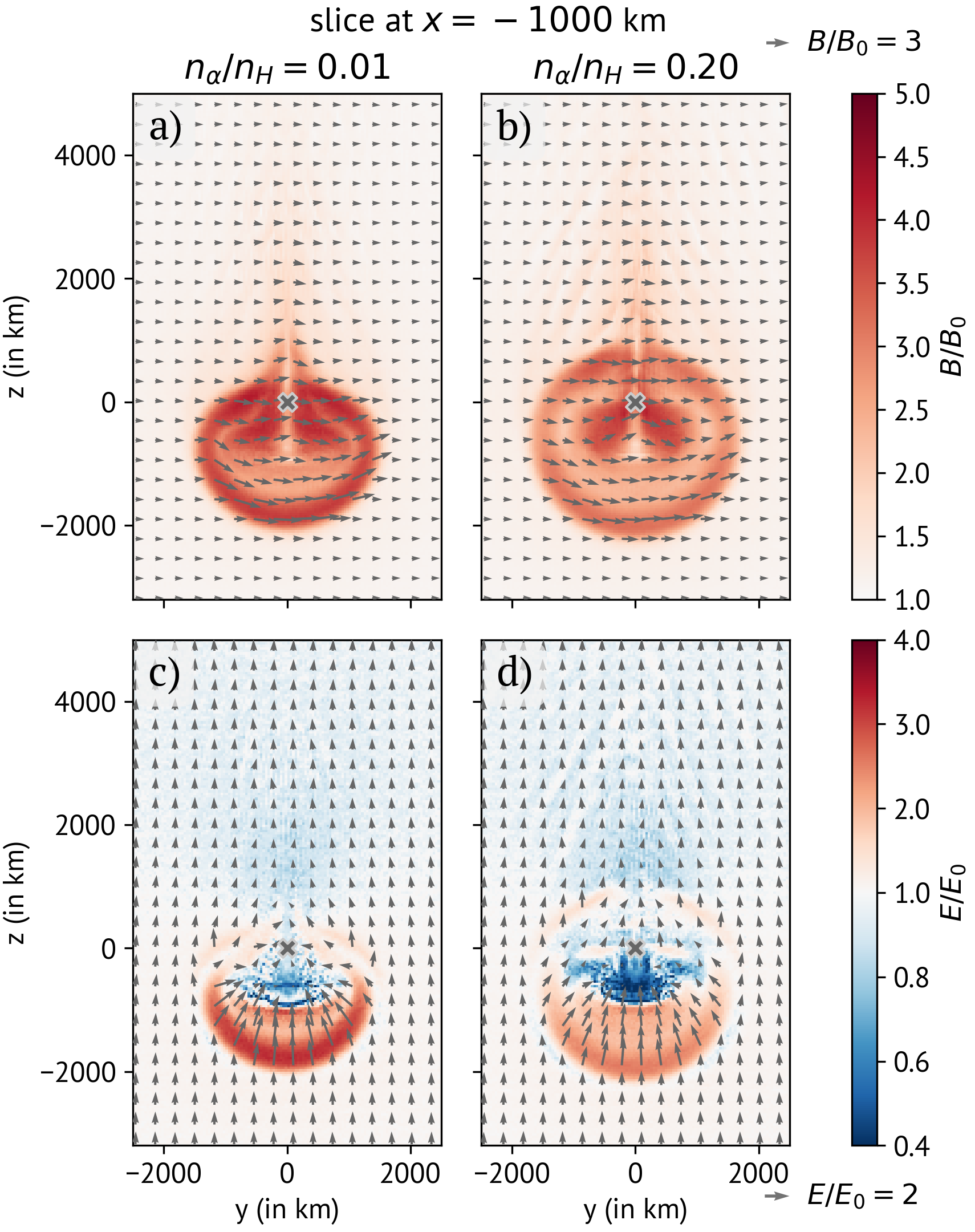}}
	\caption{Magnetic (upper row, panels a and b) and electric (lower row, panels c and d) fields in the $y-z$ plane, for the slice at $x=\SI{-1000}{km}$. The left side (panels a and c) shows the results of the low alpha simulation. The right side (panels b and d) shows the results of the high alpha simulation run. The background shows the magnitude of the electric and magnetic fields, normalised by their upstream values $B_0=\SI{3}{nT}$ and $E_0=\SI{1.29}{\mV\per m}$. The arrows show the direction of the fields in this plane, that is, $y-$ and $z-$components. The arrow length scales are specified above and below the colour bars for the respective fields.
	}
	\label{fig::E+B_x-1000}
\end{figure}

Figures \ref{fig::E+B_y0} and \ref{fig::E+B_x-1000} show a comparison between the low alpha and the high alpha simulations. The panels in Figure \ref{fig::E+B_y0} show a slice of the $x-z$ plane, at $y = \SI{0}{km}$ while the panels in Figure \ref{fig::E+B_x-1000} display a slice in the $y-z$ plane at $x= \SI{-1000}{km}$. The overall shape of the electric and magnetic fields is similar in the two cases. The magnetic field (panels a and b) piles up in front of the comet and drapes around the obstacle consisting of cometary ions. This feature is subsequently referred to as the ``magnetic field pile-up layer'' and has been observed and modelled in connection with the infant bow shock at comet 67P \citep{gunell2018aa}. 
The maximum magnetic field strength is in the $+E$-hemisphere right in front of the nucleus and is approximately the same for both runs ($\num{6.8}B_0$, where $B_0$ is the magnetic field in the upstream solar wind). In the $-E$-hemisphere (at $y=0$)  a layer with a steep increase of the magnetic field compared to the upstream value is created. As is seen in Figure \ref{fig::E+B_x-1000}, in the $y-z$ plane this layer is a circular structure reaching up into the $+E$-hemisphere, but with a gap at $y=\SI{0}{km}$ where the cometary pickup ions dominate. In the high alpha run the layer extends about \SI{200}{km} further upstream. In the low alpha case it appears compressed in the $+E$-hemisphere, that is, forming an oval rather than a circle. The enhancement of the magnetic field pile-up layer is stronger in the low alpha case ($4B_0$) compared to the high alpha case ($\num{3.4}B_0$). 

The electric field (panels c and d in Figures \ref{fig::E+B_y0} and \ref{fig::E+B_x-1000}) is also strongly enhanced in the $-E$-hemisphere at the location of the magnetic field pile-up layer. The enhancement is stronger in the low alpha case ($\num{3.3}E_0$) compared to the high alpha case ($\num{2.6}E_0$). The electric field enhancement is only observed for the lower half ($-E$ hemisphere) of the magnetic field pile-up layer, forming approximately a semi-circle in the $y-z$ plane. The direction of the electric field remains perpendicular to the layer, pointing inward towards the downstream tail. In the downstream tail region behind the nucleus the electric field is significantly decreased compared to the upstream SW. This effect is stronger in the high alpha case. The downstream $+E$-hemisphere is characterised by a slight decrease of the electric field to about \num{0.8} of its upstream value, along with a rotation of the electric field direction towards $-x$.

\subsection{Energy Transfer}
\begin{figure*}
	\centering
	\includegraphics[width=17cm]{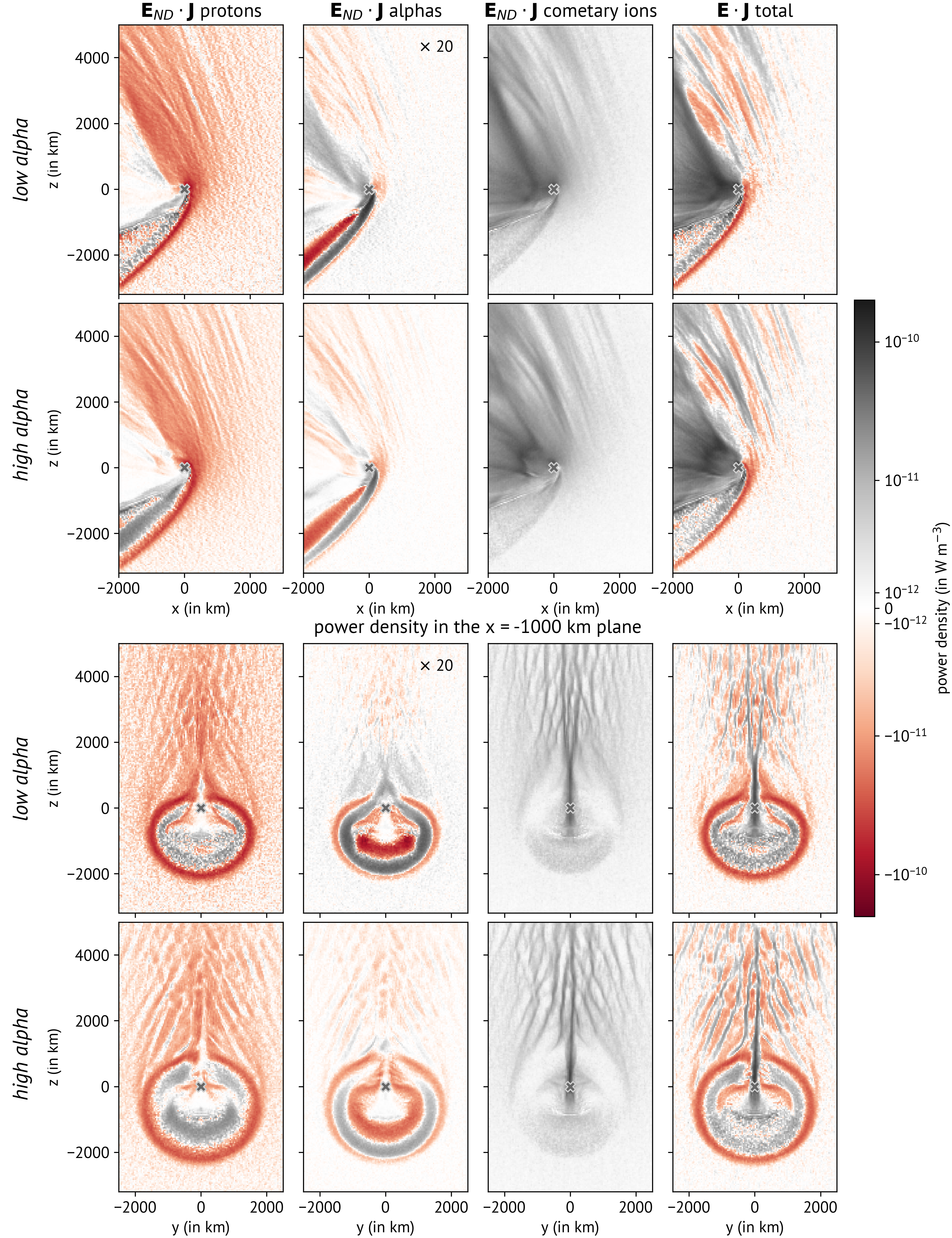}
	\caption{Power density $\mathbf{ E \cdot J }$ for the low alpha (rows 1 and 3), and the high alpha (rows 2 and 4) case. The upper two rows show the $x-z$ plane at $y=\SI{0}{km}$ (same as Figure \ref{fig::E+B_y0}), while the lower two rows show the $y-z$ plane at $x=\SI{-1000}{km}$ (same as Figure \ref{fig::E+B_x-1000}). Columns 1-3 show the power density of protons, alpha particles, and cometary ions, respectively. The last column shows the total power density $ \mathbf{E \cdot J} $. For better visibility, the results of the power density of alphas in the low alpha case (column 2, rows 1 \& 3) are scaled up by a factor of 20. }
	\label{fig::energy_transfer_planes}
\end{figure*}

\begin{figure}
	\resizebox{\hsize}{!}{\includegraphics{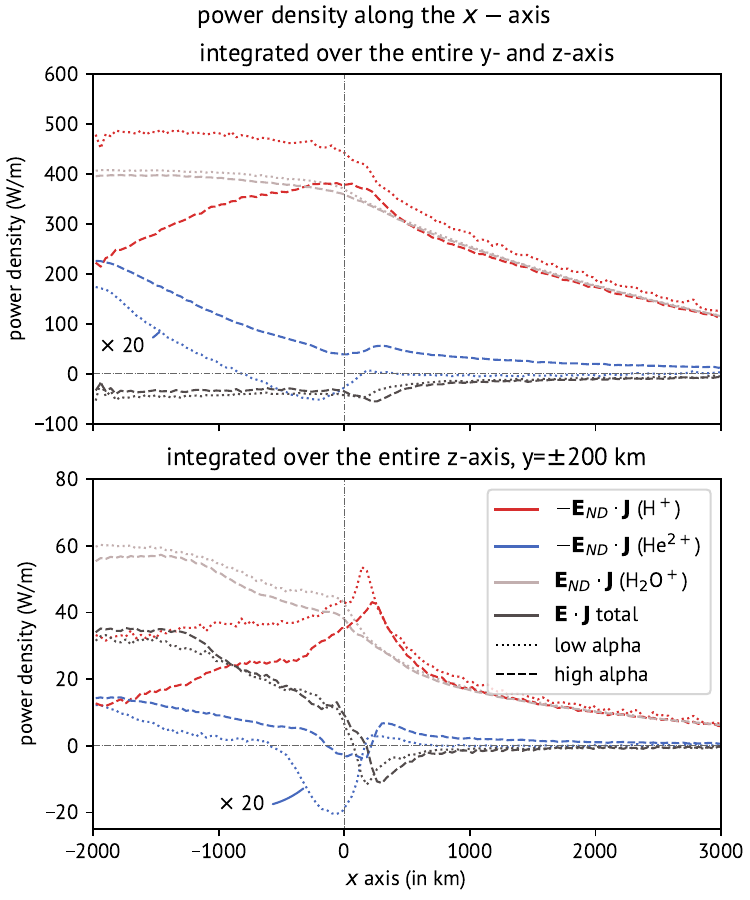}}
	\caption{Power density of the individual ion species (protons: red, alphas: blue, and cometary ions: grey) and total $ \mathbf{E \cdot J} $ (black) along the $x-$axis. The dotted and dashed lines show the results of the low and high alpha case. In the upper panel the power density is integrated along the entire $y-$ and $z-$axes. For the lower panel we only integrate between $y=\pm\SI{200}{km}$ along the $y-$axis. For the SW species the negative values are shown since these species mostly function as generators. The curve for the power density of alpha particles for the low alpha case (dotted blue line) is scaled by a factor of 20 to improve visibility. }
	\label{fig::lin_energy_transfer}
\end{figure}

In order to understand the consequences of a change in the SW alpha-to-proton ratio we focus on the energy transfer from the SW to the cometary plasma. We analyse the energy transfer in three different ways: firstly, the power density for each ion species ($\mathbf{E_{ND} \cdot J}$), as well as the total power density $(\mathbf{E \cdot J})$, integrated over the entire simulation domain, is summarised in Table \ref{tab::energy_transfer}. Secondly, Figure \ref{fig::energy_transfer_planes} shows the power density in a plane for two different slices, $y=\SI{0}{km}$, and $x=\SI{-1000}{km}$. Finally, Figure \ref{fig::lin_energy_transfer} shows the evolution of the energy transfer between the different particle populations and the fields along the $x-$axis, moving from the upstream SW to the tail region downstream.

\begin{table}[hbt]
	\caption{Integrated power density of the individual ion species, magnetic field, and total in the simulation box at time $t=\SI{118.4}{\s}$.}
	\label{tab::energy_transfer}
	\centering
  	\begin{tabular}{l l l}
   	 	& low alpha & high alpha \\
    	\hline
    	$\mathbf{E \cdot J}$ (total) & \SI{-139}{\MW} & \SI{-134}{\MW} \\
    	$\mathbf{E}_{ND} \cdot \mathbf{J}$ (H$^+$) & \SI{-1830}{\MW} & \SI{-1430}{\MW}  \\
    	$\mathbf{E}_{ND} \cdot \mathbf{J}$ (He$^{2+}$) & \SI{-4.16}{\MW} & \SI{-358}{\MW}  \\
    	$\mathbf{E}_{ND} \cdot \mathbf{J}$ (H$_2$O$^+$) & \SI{1590}{\MW} & \SI{1560}{\MW} \\
    	$\frac{\partial}{\partial t}\left( \frac{B^2}{2\mu_0} \right)$ & \SI{-4.22}{\MW} & \SI{-1.87}{\MW}
  	\end{tabular}
\end{table}

Integrated over the entire simulation domain, cometary ions act as a load. The power transfer to the cometary ions ($\mathbf{E_{ND} \cdot J}$ (H$_2$O$^+$)) is comparable for both simulation runs (cf. Table \ref{tab::energy_transfer}). The slightly higher value in the low alpha case was consistently found for three individual time steps and is expected to be a real feature, and not transient or due to numerical issues. Similarly, the total $\mathbf{E \cdot J}$ is slightly more negative in the low alpha case. In the case of protons, the power transfer is much larger in the low alpha case due to the higher number density of protons in the upstream solar wind. The opposite is true for the alpha particles. 
The change in electromagnetic energy is small, and the total $\mathbf{E\cdot J}$ is compensated by the divergence of the Poynting flux (data not shown). 

The protons act as a generator in the entire upstream region (see Figure \ref{fig::energy_transfer_planes} column 1, rows 1-2). The strongest power density is found in the initial magnetic field pile-up layer (upstream-facing part). Behind the initial pile-up the protons gain energy, as seen by the positive power density. This feature is more pronounced in the high alpha case. In the $+E$-hemisphere the power density of the protons is mostly negative and appears modulated by wave-like structures. The alphas (column 2) show similar structures as the protons, but the initial generator region is much weaker.  
Compared to the protons there are more regions where the alphas act as a load, including large parts of the $+E$-hemisphere. Upstream of the nucleus the power density of alphas is close to zero, or even slightly positive in the low alpha case. Comparing the low and high alpha case (rows 1\&3 vs. 2\&4 in Figure \ref{fig::energy_transfer_planes}) shows that the structures are similar but the magnitude of the power density of alphas is much smaller in the low alpha case due to the significantly lower number densities. The cometary ions (column 3) are a load in the entire simulation domain ($\mathbf{E}_{ND} \cdot \mathbf{J}>0$). Most of the energy transfer is concentrated near $y=0$ in the $+E$-hemisphere and the tail region. The power density outside the plane at $y=0$, as well as in the upstream region, appears in a wave-like pattern. The total $\mathbf{E \cdot J}$ (right column, Figure \ref{fig::energy_transfer_planes}) is dominated by SW protons and cometary ions. The effect of alphas is only visible in the high alpha case, especially in areas where the power density of cometary ions is fairly low (for example at $y=\pm\SI{1000}{km}$, between $z=0$ and $z=\SI{-800}{km}$, downstream of the magnetic field pile-up layer).

Figure \ref{fig::lin_energy_transfer} shows two views of the evolution of the power density along $x$. In the upper panel we integrate the power density over $y$ and $z$ using the full range of the simulation domain, therefore capturing the energy transfer in the entire simulation box. In the lower panel only the power density within $y=\pm\SI{200}{km}$ is included in the integration and the result is more comparable with the $x-z$ view shown in Figure \ref{fig::energy_transfer_planes} in the upper two rows. The power density of both solar wind species are shown with the opposite sign $(-\mathbf{ E_\mathrm{ND} \cdot J })$ for easier comparison with the cometary ions. 

Upstream $(x \gg 0)$ the power density approaches zero with increasing $x$. There is a net transfer of energy from the SW ions to the cometary ions and the electromagnetic fields. The power density of protons is much higher compared to the alphas, even in the high alpha case where their mass densities are almost comparable. In the downstream region ($x<0$) the energy transfer to the cometary ions as well as the total $\mathbf{ E \cdot J }$ integrated over the entire $y$-axis remains almost constant. This is also the case for the protons in the low alpha simulation run. In this case the alpha particles act as a load from \SI{200}{km} upstream of the nucleus to \SI{800}{km} downstream. In the region further downstream the alpha population acts as a generator, although its contribution to the total energy transfer is very small compared to the other ion species. For the high alpha simulation the absolute value of the power density of protons reaches a peak of \SI{400}{W\per\m} around $x=0$ and decreases in magnitude further downstream. The power density of alphas on the other hand starts to increase in magnitude downstream of the nucleus, reaching a value of about \SI{-230}{W\per\m} at the downstream boundary, comparable to the power density of protons at this point.

The power density close to $y=0$ (Figure \ref{fig::lin_energy_transfer}, lower panel) exhibits a clear peak in the absolute value of the power density of protons for both the low and high alpha case (\SI{53}{W\per\m} and \SI{41}{W\per\m}, respectively). The peak in the high alpha case is located approximately \SI{150}{km} further upstream, consistent with the slightly larger magnetosphere seen in Figure \ref{fig::E+B_y0}). In the downstream region the power density of protons in the low alpha case is twice that of the high alpha case. The power density of alpha particles in both cases have a corresponding peak in approximately the same region as the protons. Moving downstream they transition into a load region (power density $>0$). Even further downstream the power density becomes negative again. The power density of cometary ions shows the steepest increase at $x=0$, and remains approximately constant after $x=\SI{-1200}{km}$. The total $\mathbf{ E \cdot J }$ has a local minimum at the same location as the proton power density, and the values are similar for both simulation runs. Moving further downstream, the total $\mathbf{ E \cdot J }$ quickly becomes positive, contrary to the results of the full simulation (upper panel). This indicates a transfer of electromagnetic field energy along the $y$-axis out of the slice around $y=\pm\SI{200}{km}$. 

\section{Discussion}

The bulk properties of the solar wind such as speed, density, and the resulting dynamic pressure are well known to affect its interaction with solar system bodies. For higher pressure the boundaries, such as the bow shock and magnetopause, are found closer to the object \citep{shue_2002, edberg_mars_2010}. At Earth a higher solar wind dynamic pressure leads to higher ion escape rates \citep{schillings_2019}. At Mars a weak inverse relation between solar wind dynamic pressure and ion escape has been found \citep{ramstad_2018}. In our simulation study we have looked at the effect of varying composition of the solar wind for a fixed dynamic pressure. Our results show that, even for a low-activity comet, the composition of the solar wind influences the spatial structure and the manner in which energy is transferred from the solar wind to the cometary ions.

\subsection{Spatial Structure}

The initial formation of the magnetic pile-up layer in the $-E$ hemisphere is mostly driven by the protons.
The electron fluid in the model is mass less and charge neutralising so it will immediately follow the ions. 
The magnetic field in turn is frozen to the electron fluid in our model. In the region dominated by the solar wind ions, i.\,e. the $-E$ hemisphere, the magnetic field strength distribution will thus be determined by the solar wind ion charge density distribution. Due to different gyroradii, the locations of the density enhancement of protons and alphas are different.
In the high alpha case the region with enhanced solar wind charge density is broader since the alphas contribute significantly to the total charge density. The relative enhancement with respect to the upstream SW density, and thus the magnetic field strength in the magnetic field pile-up layer, is therefore lower compared to the low alpha case (not shown).
One may expect a second magnetic pile-up associated with the alpha particles in the high alpha case. This is not seen in our simulations. The reason appears to be that a magnetic field enhancement has already been caused by the protons and these still dominate in number and charge density, even though the alpha particles carry a similar amount of kinetic energy / momentum flux. 

Why the magnetic field pile-up and proton density enhancement layer in the $-E$ hemisphere occur further upstream for the high alpha case can be explained in a similar manner. 
The amount of mass-loading (cometary ions) is the same as in the low alpha case but the mass density of protons is smaller. Thus there is relatively more mass-loading of the proton population and thus the boundaries form further upstream. 
In the high alpha case the alpha particles gradually gain importance further downstream from the magnetic field pile-up layer as discussed below.

\subsection{Energy Transfer}

The energy transfer from the SW to the cometary ions in the upstream region is much more efficient for the protons than the alpha particles. This is a direct consequence of the difference in deflection due to their different gyroradii.
As the SW is mass-loaded by the cometary plasma, its individual components react in different ways: the cometary ions are accelerated along the electric field towards $+z$. To conserve momentum, the SW ions are deflected towards $-z$, which can also be seen as a rotation/gyration of $\mathbf{v}_\mathrm{SW}$ around the magnetic field.  
Upstream of the magnetic field pile-up layer the electric current density $\mathbf{J}$ is negligible (cf. Eq. \ref{eq::J_hybrid}), and the electron velocity $\mathbf{v}_e$ is approximately the ion velocity $\mathbf{v}_i \approx \mathbf{v}_e$, where $\mathbf{v}_\mathrm{i} = \mathbf{J}_i/\rho_i$ $(\mathbf{J}_i \approx -\mathbf{J}_e,\;\rho_i=-\rho_e)$. 
The cometary ions are accelerated along the electric field, and the net ion velocity $\mathbf{v}_i$, and therefore $\mathbf{v}_e$, differs from the SW flow $\mathbf{v}_\mathrm{SW}$.
As a result, the electric field, carried by the electrons (see Eq. \ref{eq::E_ND_hybrid}), has a component directed against the solar wind velocity.
Therefore, the angle between $\vec{v}_\mathrm{SW}$ and the electric field, $ \angle(\vec{v}_\mathrm{SW}, \vec{E})$ is $ >\ang{90}$, which also corresponds to a negative $\mathbf{E_\mathrm{ND} \cdot J}$ of the solar wind ions, i.\,e.,  a generator region. Equivalently one can look at the flow direction of solar wind ions with respect to the electrons. A solar wind flow clock-wise of the electron flow indicates a load.

In case of a two-species SW the SW velocity depends on the velocity and charge density of each individual species. 
If one SW species dominates the SW mass density, like the protons in our low alpha case, the SW velocity will approximately correspond to the bulk velocity of this major species $( \vec{v}_\mathrm{SW} \approx \vec{v}_\mathrm{bulk, major} )$. The deflection of the second, minor species can be approximated as single particle motion in the given electromagnetic fields and depends on its inertia (mass-per-charge $m/q$). In a typical scenario the major SW species will be protons, which have the lowest $m/q$ of all ions and therefore the highest deflection. The minor species with a larger $m/q$ will be less deflected, and $ \angle(\vec{v}_\mathrm{bulk, minor}, \vec{E}) < \angle(\vec{v}_\mathrm{bulk, major}, \vec{E})$. Depending on the $m/q$ of the minor species and the electromagnetic fields this angle can be above or below \ang{90}, leading to a possible case where part of the solar wind acts as a load ($\mathbf{E \cdot J}$ > 0 for the minor species). 
This can primarily be seen during initial deflection, before particle trajectories get too complex, and is stronger for ions with larger $m/q$. As the mass density of the minor species increases, $ \vec{v}_\mathrm{SW} $ and the related electric field direction are more influenced by this species, and $ \angle(\vec{v}_\mathrm{bulk, minor}, \vec{E}) $ will drop below \ang{90}, ensuring that this species acts as a generator. Our results (Figs. \ref{fig::energy_transfer_planes} and \ref{fig::lin_energy_transfer}) show this effect up to \SI{500}{km} upstream of the nucleus: in the low alpha case the power density of alphas is weakly positive in some regions, and in the high alpha case the power density is negative but much smaller in magnitude compared to the protons, despite having \SI{40}{\percent} of the proton's charge density (and therefore current). 

\begin{figure*}
	\sidecaption
	\includegraphics[width=12cm]{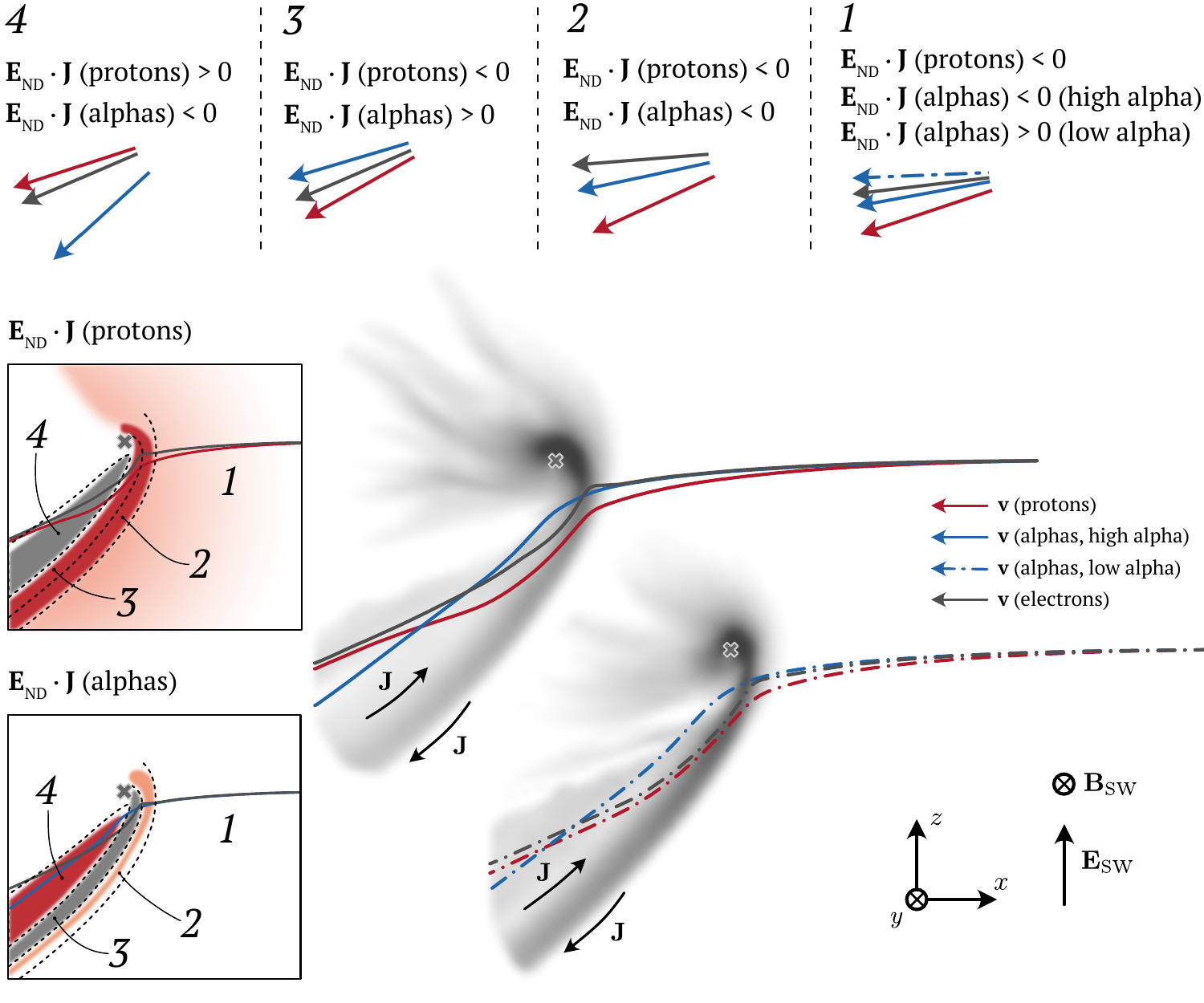}
	\caption{Illustration of proton, alpha, and electron bulk velocities and derived flow lines in the $x-z$ plane at $y=\SI{0}{km}$. The upper row of arrows show the plasma velocities in different areas (marked 1~--~4), corresponding to different combinations of generator/load regions for alphas and protons. The deflection angles are exaggerated for better illustration. The areas are defined in the two insets to the right, which show the $\mathbf{ E_\mathrm{ND} \cdot J }$ for protons and alphas in the high alpha case (cf. Fig. \ref{fig::energy_transfer_planes}, row 2). The lower right part of the figure shows the flow lines for the low and high alpha case, with the magnitude of the magnetic field (cf. Fig. \ref{fig::E+B_y0}, panels a and b) in greyscale in the background. There the relative position of the flow lines with respect to the magnetic field pile-up layer can be seen.
	Solid lines: high alpha case, dash-dotted lines: low alpha case. Red, blue, and grey indicate protons, alphas, and electrons, respectively. The grey `x' marker indicates the position of the comet nucleus.
	The flow line plots for the low and high alpha case are separated for improved visibility. }
	\label{fig::flowlines}
\end{figure*}

Figure \ref{fig::flowlines} illustrates this effect using $\mathbf{v}_e$ instead of $\mathbf{E}$. This makes it easier to visualise generator and load regions based on the flow direction of ions and electrons. A region where the ions are rotated clockwise with respect to the electrons is a load region for this ion species.
We identified four regions determined by the sign of the power density of protons and alphas, defined as area `1' -- `4' in the insets to the left (based on Figure 3).
 The illustrated flow lines are calculated from the simulated bulk velocities of protons, alphas, and electrons (Note: these flow lines do not necessarily correspond to particle trajectories in case of the ions if the velocity distributions are not Maxwellian). 
The upstream case described in the previous paragraph corresponds to area `1' (arrows in the top left). The blue arrows corresponding to the alphas may rotate slower (low alpha, dash-dotted line) or faster (high alpha, solid line) than $\mathbf{v}_e$. In our high alpha simulation the alpha bulk velocity is almost aligned with $\mathbf{v}_e$, leading to a negative, but very small (in magnitude) power density. For an efficient energy transfer of alphas to the cometary ions in this region an even higher alpha-to-proton ratio would be required.

Close to the pile-up layer (area `2') the flow direction of the protons and, to a lesser extent, the alphas, changes more suddenly. The $E-$field and electron velocity also make a sudden change due to the currents (see $\mathbf{J}$ arrows in Fig. \ref{fig::flowlines}).
The net result is a clockwise rotation of $\mathbf{v}_e$, also seen as a small bump towards $+z$ in the electron flow line of the high alpha case in Fig. \ref{fig::flowlines}. As a consequence both alphas and protons act as generators in this area. As soon as the current density decreases in magnitude and changes sign downstream of the magnetic field pile-up layer the electric field rotates counter-clockwise again, resulting in $\mathbf{E_\mathrm{ND} \cdot J} > 0$ for the alphas (area `3'). The structure of this load-generator region near the pile-up layer appears to be similar for both high and low alpha ratio cases.

Further downstream (area `4' in Fig. \ref{fig::flowlines}) the electric field changes less, and the alphas gyrate further, resulting in a negative power density, or a generator region.
The deflection of protons decreases as their gyration progresses, and they form a load region downstream of the pile-up layer.
This feature is seen in both cases and has been previously reported in \citet{lindkvist_energy_2018}. They simulate a comet with higher outgassing rate and only include one SW species (protons). Their results show several transitions between generator and load regions in the power density of protons. These striations in the SW proton power density are attributed to the secondary acceleration of protons due to a change in current direction.
Compared with our low alpha case (Figure \ref{fig::energy_transfer_planes}, rows 1\&3), there are more generator-load transitions in the protons in \citet{lindkvist_energy_2018} due to the broader pile-up layer. The transitions between negative and positive power density, especially the first load region in the protons, are more distinct in our high alpha case. This is likely due to the two SW species mass-loading each other and affecting the ion dynamics of the other species.

Every cometary ion which is created will eventually be picked up and reach solar wind speed. Hence, the total transfer of solar wind energy to the cometary ions must be the same regardless of the solar wind alpha-to-proton ratio as long as the newborn ion profiles are identical. Since there can be spatial and temporal variations in the simulations the power density must be integrated over a sufficiently large volume and averaged in time to cancel out non-negligible fluctuations. Far downstream of the nucleus the cometary ion production is negligible. The number flux of cometary ions is then conserved, that is, constant through all slices $x = x_{S}$ of the simulation box, for $x_S $ sufficiently far downstream. From Figure~\ref{fig::E+B_y0}c and d we note that the electric field in the region dominated by cometary ions -- downstream of the nucleus and behind the pile-up layer -- is not varying much along x.  If we assume the field to remain constant along the flow lines of the cometary ions in this region (within the domain covered by the simulation box), the power density of the cometary ions will also be constant if the flux is conserved. This is indeed what we see in the upper panel of Figure~\ref{fig::lin_energy_transfer}: Downstream of $x \approx \SI{-1000}{km}$, $\mathbf{E}_{ND} \cdot \mathbf{J}$ for the cometary ions is almost constant for both runs.

The energy transfer from the SW to the cometary ions is affected by the amount of alphas present. In the upper panel of Figure~\ref{fig::lin_energy_transfer} a clear difference is noted between the high and low alpha runs, especially downstream of the nucleus ($x<0$). For the high alpha run, the decrease in the integrated power density ($\mathbf{\int_{YZ}E_{ND} \cdot J}$) of the protons is compensated by the increase of the power density of the alphas. The two solar wind species are coupled to each other and their interaction modifies the spatial distribution of the power density. 

In the high alpha case (where the alpha particles constitute $20\%$ of the total number density) the kinetic energy available in the SW alpha particles is $80\%$ of the energy available in the protons (as the mass of an alpha particle is four times the mass of a proton and the velocities are equal). Still the power density of the solar wind species only reach the same value at the left edge of the simulation box. Within this box the power transfer from the alphas to the cometary ions and fields is much less than that from the protons (cf. Table~\ref{tab::energy_transfer}). This is likely caused by the delayed mass-loading of the alpha particles.

In the spatial domain covered in the simulations the distance from the pile-up layer to the sun-comet line increases almost linearly with a decreasing $x$ (cf. Figure~\ref{fig::E+B_y0}). Provided the magnetic field strength and the thickness of the pile-up layer does not change much the magnetic energy density contained in the layer must increase linearly with $-x$, as the radius of the almost circular layer increases (cf. Figure~\ref{fig::E+B_x-1000}). This means that the power transfer to the fields must be constant, which is what we see for $x<0$ in Figure~\ref{fig::lin_energy_transfer} (upper panel). Close to the nose of the cometary magnetosphere, where there is a significant curvature of the pile-up layer, this will not be true and we observe a peak in the magnitude of the power density.

With both the total $\mathbf{E \cdot J}$ and power density of the cometary ions being approximately constant downstream of the nucleus, it follows that the sum of the power densities for protons and alpha particles must be constant. In the high alpha case it means that when the power density of the alpha population is increasing in this region, there is a corresponding decrease of the power density carried by the protons. 

It should also be noted that the magnitude of the total power density is considerably smaller than the power density of the particle populations, with exception of the alpha particles in the low alpha case. This suggests that mass-loading is the dominant process of energy transfer between SW and cometary ions.

\section{Conclusions}
In this paper we used a hybrid model to investigate the effect of different SW alpha-to-proton ratios on the solar wind's interaction with a comet. We made two simulation runs; one run with the ratio between the number densities of alphas and protons being 1\% and another run with a ratio of 20\%. The number densities are adjusted so that the dynamic pressure and momentum of the SW are the same in the two cases. 

By comparing the two runs we conclude that a higher SW alpha-to-proton ratio leads to a larger cometary magnetosphere. The initial pile-up of the magnetic field is governed by the protons also in the high alpha case. With less SW protons available the relative mass loading of the SW protons is larger, pushing the pile-up layer further away from the nucleus. The electromagnetic fields in the pile-up layer are weaker in this case as the fraction of the total magnetic field frozen into the electron fluid related to the SW protons is smaller.  

The interaction between the cometary atmosphere and the solar wind leads to a transfer of energy from the SW species to the cometary ions. The transfer of energy from the alpha particles to the cometary ions occurs further downstream than for the protons due to the larger inertia of the alpha particles. As a result, there are regions downstream where the power density of alphas is comparable to the power density of protons. 
The two SW species take turns in acting as loads and generators. In well-defined spatial regions the sign of the power density is different for protons and the alpha particles. The SW species gain energy in some regions as a result of the alpha and proton populations mass loading each other.

If the SW is comprised of multiple ion species, mass loading and energy transfer to the cometary ions will be most efficient for the species with lowest $m/q$, typically protons. This is due to the different deflection of the individual species, caused by the different gyroradii. In some situations it is possible that the SW species with higher inertia has a positive power density, i.\,e., acts as a load, in some regions.
When analysing the impacts of CMEs on comets or other unmagnetised bodies, the effects of a potentially increased alpha-to-proton ratio onto the induced magnetospheres shown in this paper should be taken into account.

\section*{Acknowledgements}
The simulations were enabled by resources provided by the EuroHPC Regular Access at Vega, grant EHPC-REG-2023R03-023, and by the National Academic Infrastructure for Supercomputing in Sweden (NAISS) at High Performance Computing Center North (HPC2N) partially funded by the Swedish Research Council through grant agreement no. 2022-06725.
Work at the Swedish Institute of Space Physics in Kiruna (IRF) was funded by the Swedish National Space Agency (SNSA) grant 132/19.
Work at Ume\aa\ university was supported by SNSA grant 2023-00208 (HG) and SNSA grant 2022-00183 (SF).

\bibliographystyle{bibtex/aa} 
\bibliography{bibtex/references.bib}

\begin{thebibliography}{46}
\expandafter\ifx\csname natexlab\endcsname\relax\def\natexlab#1{#1}\fi

\bibitem[{Aizawa {et~al.}(2021)Aizawa, Griton, Fatemi, Exner, Deca, Pantellini,
  Yagi, Heyner, Génot, André, Amaya, Murakami, Beigbeder, Gangloff,
  Bouchemit, Budnik, \& Usui}]{aizawa_cross-comparison_2021}
Aizawa, S., Griton, L., Fatemi, S., {et~al.} 2021, Planetary and Space Science,
  198, 105176

\bibitem[{Alfv{\'e}n(1957)}]{alfven1957tellus}
Alfv{\'e}n, H. 1957, Tellus, IX, 92

\bibitem[{{Behar} {et~al.}(2016){Behar}, {Lindkvist}, {Nilsson},
  {Holmstr{\"o}m}, {Stenberg-Wieser}, {Ramstad}, \& {G{\"o}tz}}]{behar2016aa}
{Behar}, E., {Lindkvist}, J., {Nilsson}, H., {et~al.} 2016, Astronomy and
  Astrophysics, 596, A42

\bibitem[{{Behar} {et~al.}(2018){Behar}, Tabone, Saillenfest, Deca,
  Holmstr\"{o}m, \& Nilsson}]{behar2018baa}
{Behar}, E., Tabone, B., Saillenfest, M., {et~al.} 2018, Astron. Astrophys.,
  620, A35

\bibitem[{Biermann(1951)}]{biermann1951za}
Biermann, L. 1951, Zeitschrift fur Astrophysik, 29, 274

\bibitem[{Cravens \& Gombosi(2004)}]{cravens2004asr}
Cravens, T. \& Gombosi, T. 2004, Advances in Space Research, 33, 1968 ,
  {Comparative Magnetospheres}

\bibitem[{{Edberg} {et~al.}(2016){Edberg}, {Alho}, {Andr{\'e}}, {Andrews},
  {Behar}, {Burch}, {Carr}, {Cupido}, {Engelhardt}, {Eriksson}, {Glassmeier},
  {Goetz}, {Goldstein}, {Henri}, {Johansson}, {Koenders}, {Mandt}, {M{\"o}stl},
  {Nilsson}, {Odelstad}, {Richter}, {Simon Wedlund}, {Stenberg Wieser},
  {Szego}, {Vigren}, \& {Volwerk}}]{edberg2016mnras}
{Edberg}, N.~J.~T., {Alho}, M., {Andr{\'e}}, M., {et~al.} 2016, Monthly Not.
  Roy. Astr. Soc., 462, S45

\bibitem[{Edberg {et~al.}(2010)Edberg, Lester, Cowley, Brain, Fränz, \&
  Barabash}]{edberg_mars_2010}
Edberg, N. J.~T., Lester, M., Cowley, S. W.~H., {et~al.} 2010, Journal of
  Geophysical Research: Space Physics, 115
  [\eprint{https://agupubs.onlinelibrary.wiley.com/doi/pdf/10.1029/2009JA014998}]

\bibitem[{Fatemi {et~al.}(2024)Fatemi, Hamrin, Krämer, Gunell, Nordin,
  Karlsson, \& Goncharov}]{fatemi_jets_2024}
Fatemi, S., Hamrin, M., Krämer, E., {et~al.} 2024, Monthly Notices of the
  Royal Astronomical Society, 531, 4692

\bibitem[{Fatemi {et~al.}(2018)Fatemi, Poirier, Holmström, Lindkvist, Wieser,
  \& Barabash}]{fatemi_mercury_2018}
Fatemi, S., Poirier, N., Holmström, M., {et~al.} 2018, Astronomy \&
  Astrophysics, 614, A132

\bibitem[{Fatemi {et~al.}(2017)Fatemi, Poppe, Delory, \&
  Farrell}]{fatemi_amitis_2017}
Fatemi, S., Poppe, A.~R., Delory, G.~T., \& Farrell, W.~M. 2017, in Journal of
  {Physics}: {Conference} {Series}, Vol. 837 (Institute of Physics Publishing)

\bibitem[{Fatemi {et~al.}(2022)Fatemi, Poppe, Vorburger, Lindkvist, \&
  Hamrin}]{fatemi_ganymede_2022}
Fatemi, S., Poppe, A.~R., Vorburger, A., Lindkvist, J., \& Hamrin, M. 2022,
  Journal of Geophysical Research: Space Physics, 127, e2021JA029863

\bibitem[{Glassmeier {et~al.}(2007)Glassmeier, Boehnhardt, Koschny, K{\"u}hrt,
  \& Richter}]{glassmeier2007ssr}
Glassmeier, K.-H., Boehnhardt, H., Koschny, D., K{\"u}hrt, E., \& Richter, I.
  2007, Space Science Reviews, 128, 1

\bibitem[{Goetz {et~al.}(2022)Goetz, Behar, Beth, Bodewits, Bromley, Burch,
  Deca, Divin, Eriksson, Feldman, Galand, Gunell, Henri, Heritier, Jones,
  Mandt, Nilsson, Noonan, Odelstad, Parker, Rubin, Simon~Wedlund, Stephenson,
  Taylor, Vigren, Vines, \& Volwerk}]{goetz2022ssr}
Goetz, C., Behar, E., Beth, A., {et~al.} 2022, Space Science Reviews, 218, 65

\bibitem[{Gulkis {et~al.}(2015)Gulkis, Allen, Von~Allmen, Beaudin, Biver,
  Bockelée-Morvan, Choukroun, Crovisier, Davidsson, Encrenaz, Encrenaz,
  Frerking, Hartogh, Hofstadter, Ip, Janssen, Jarchow, Keihm, Lee, Lellouch,
  Leyrat, Rezac, Schloerb, \& Spilker}]{gulkis_subsurface_2015}
Gulkis, S., Allen, M., Von~Allmen, P., {et~al.} 2015, Science, 347, aaa0709

\bibitem[{Gunell {et~al.}(2024)Gunell, Goetz, \& Fatemi}]{gunell_radial_2024}
Gunell, H., Goetz, C., \& Fatemi, S. 2024, Astronomy \& Astrophysics, 682, A62

\bibitem[{Gunell {et~al.}(2018)Gunell, Goetz, Simon~Wedlund, Lindkvist, Hamrin,
  Nilsson, Llera, Eriksson, \& Holmstr{\"o}m}]{gunell2018aa}
Gunell, H., Goetz, C., Simon~Wedlund, C., {et~al.} 2018, Astronomy and
  Astrophysics, 619, L2

\bibitem[{Hamrin {et~al.}(2012)Hamrin, Marghitu, Norqvist, Buchert, Andr{\'e},
  Klecker, Kistler, \& Dandouras}]{hamrin2012jgr}
Hamrin, M., Marghitu, O., Norqvist, P., {et~al.} 2012, Journal of Geophysical
  Research: Space Physics, 117
  [\eprint{https://agupubs.onlinelibrary.wiley.com/doi/pdf/10.1029/2012JA017707}]

\bibitem[{Hamrin {et~al.}(2006)Hamrin, Marghitu, R\"onnmark, Klecker, Andr\'e,
  Buchert, Kistler, McFadden, R\`eme, \& Vaivads}]{hamrin2006ag}
Hamrin, M., Marghitu, O., R\"onnmark, K., {et~al.} 2006, Annales Geophysicae,
  24, 637

\bibitem[{Hansen {et~al.}(2016)Hansen, Altwegg, Berthelier, Bieler, Biver,
  Bockelée-Morvan, Calmonte, Capaccioni, Combi, De~Keyser, Fiethe, Fougere,
  Fuselier, Gasc, Gombosi, Huang, Le~Roy, Lee, Nilsson, Rubin, Shou, Snodgrass,
  Tenishev, Toth, Tzou, \& Simon~Wedlund}]{hansen_evolution_2016}
Hansen, K.~C., Altwegg, K., Berthelier, J.~J., {et~al.} 2016, Monthly Notices
  of the Royal Astronomical Society, 462, S491, publisher: Oxford University
  Press

\bibitem[{Haser(1957)}]{haser_1957}
Haser, L. 1957, Bulletin de la Societe Royale des Sciences de Liege, 43, 740

\bibitem[{Heritier {et~al.}(2018)Heritier, Galand, Henri, Johansson, Beth,
  Eriksson, Vallières, Altwegg, Burch, Carr, Ducrot, Hajra, \&
  Rubin}]{heritier_plasma_2018}
Heritier, K.~L., Galand, M., Henri, P., {et~al.} 2018, Astronomy \&
  Astrophysics, 618, A77

\bibitem[{Jia {et~al.}(2007)Jia, Combi, Hansen, \& Gombosi}]{jia2007jgr}
Jia, Y.-D., Combi, M.~R., Hansen, K.~C., \& Gombosi, T.~I. 2007, Journal of
  Geophysical Research: Space Physics, 112
  [\eprint{https://agupubs.onlinelibrary.wiley.com/doi/pdf/10.1029/2006JA012175}]

\bibitem[{Kasper {et~al.}(2007)Kasper, Stevens, Lazarus, Steinberg, \&
  Ogilvie}]{kasper_2007}
Kasper, J.~C., Stevens, M.~L., Lazarus, A.~J., Steinberg, J.~T., \& Ogilvie,
  K.~W. 2007, The Astrophysical Journal, 660, 901

\bibitem[{Ledvina {et~al.}(2008)Ledvina, Ma, \& Kallio}]{ledvina_modeling_2008}
Ledvina, S.~A., Ma, Y.-J., \& Kallio, E. 2008, Space Science Reviews, 139, 143

\bibitem[{Lee {et~al.}(2015)Lee, Von~Allmen, Allen, Beaudin, Biver,
  Bockelée-Morvan, Choukroun, Crovisier, Encrenaz, Frerking, Gulkis, Hartogh,
  Hofstadter, Ip, Janssen, Jarchow, Keihm, Lellouch, Leyrat, Rezac, Schloerb,
  Spilker, Gaskell, Jorda, Keller, \& Sierks}]{lee_spatial_2015}
Lee, S., Von~Allmen, P., Allen, M., {et~al.} 2015, Astronomy \& Astrophysics,
  583, A5

\bibitem[{Lindkvist {et~al.}(2018)Lindkvist, Hamrin, Gunell, Nilsson, Wedlund,
  Kallio, Mann, Pitkänen, \& Karlsson}]{lindkvist_energy_2018}
Lindkvist, J., Hamrin, M., Gunell, H., {et~al.} 2018, Astronomy \&
  Astrophysics, 616, A81

\bibitem[{Lockwood(2019)}]{lockwood2019jgr}
Lockwood, M. 2019, Journal of Geophysical Research: Space Physics, 124, 5498

\bibitem[{Marghitu {et~al.}(2006)Marghitu, Hamrin, Klecker, Vaivads, McFadden,
  Buchert, Kistler, Dandouras, Andr\'e, \& R\`eme}]{marghito2006ag}
Marghitu, O., Hamrin, M., Klecker, B., {et~al.} 2006, Annales Geophysicae, 24,
  619

\bibitem[{{Moeslinger} {et~al.}(2024){Moeslinger}, Gunell, Nilsson, Fatemi, \&
  Stenberg~Wieser}]{moeslinger2024essoar}
{Moeslinger}, A., Gunell, H., Nilsson, H., Fatemi, S., \& Stenberg~Wieser, G.
  2024, ESS Open Archive

\bibitem[{Moeslinger {et~al.}(2023{\natexlab{a}})Moeslinger, Nilsson,
  Stenberg~Wieser, Gunell, \& Goetz}]{moeslinger2023bjgr}
Moeslinger, A., Nilsson, H., Stenberg~Wieser, G., Gunell, H., \& Goetz, C.
  2023{\natexlab{a}}, Journal of Geophysical Research: Space Physics, 128,
  e2023JA031746, e2023JA031746 2023JA031746

\bibitem[{Moeslinger {et~al.}(2023{\natexlab{b}})Moeslinger, Wieser, Nilsson,
  Gunell, Williamson, LLera, Odelstad, \& Richter}]{moeslinger2023ajgr}
Moeslinger, A., Wieser, G.~S., Nilsson, H., {et~al.} 2023{\natexlab{b}},
  Journal of Geophysical Research: Space Physics, 128, e2022JA031082,
  e2022JA031082 2022JA031082

\bibitem[{{Nilsson} {et~al.}(2018){Nilsson}, {Gunell}, {Karlsson}, {Brenning},
  {Henri}, {Goetz}, {Eriksson}, {Behar}, {Stenberg Wieser}, \&
  {Valli{\`e}res}}]{nilsson2018aa}
{Nilsson}, H., {Gunell}, H., {Karlsson}, T., {et~al.} 2018, Astron. Astrophys.,
  616, A50

\bibitem[{Olsson \& Barger(1987)}]{olsson1987book}
Olsson, M.~G. \& Barger, V. 1987, Classical Electricity and Magnetism: A
  Contemporary Perspective (Allyn and Bacon)

\bibitem[{Ramstad {et~al.}(2018)Ramstad, Barabash, Futaana, Nilsson, \&
  Holmström}]{ramstad_2018}
Ramstad, R., Barabash, S., Futaana, Y., Nilsson, H., \& Holmström, M. 2018,
  Journal of Geophysical Research: Planets, 123, 3051

\bibitem[{Reisenfeld {et~al.}(2001)Reisenfeld, Gary, Gosling, Steinberg,
  McComas, Goldstein, \& Neugebauer}]{reisenfeld_helium_2001}
Reisenfeld, D.~B., Gary, S.~P., Gosling, J.~T., {et~al.} 2001, Journal of
  Geophysical Research: Space Physics, 106, 5693

\bibitem[{Saunders \& Russell(1986)}]{saunders1986jgr}
Saunders, M.~A. \& Russell, C.~T. 1986, Journal of Geophysical Research: Space
  Physics, 91, 5589

\bibitem[{{Schillings} {et~al.}(2019){Schillings}, {Slapak}, {Nilsson},
  {Yamauchi}, {Dandouras}, \& {Westerberg}}]{schillings_2019}
{Schillings}, A., {Slapak}, R., {Nilsson}, H., {et~al.} 2019, Earth, Planets
  and Space, 71, 70

\bibitem[{{Shue} \& {Song}(2002)}]{shue_2002}
{Shue}, J.~H. \& {Song}, P. 2002, Plan. Space Sci., 50, 549

\bibitem[{Song {et~al.}(2022)Song, Cheng, Li, Zhang, \& Chen}]{song_2022}
Song, H., Cheng, X., Li, L., Zhang, J., \& Chen, Y. 2022, The Astrophysical
  Journal, 925, 137

\bibitem[{Vourlidas {et~al.}(2007)Vourlidas, Davis, Eyles, Crothers, Harrison,
  Howard, Moses, \& Socker}]{vourlidas2007apj}
Vourlidas, A., Davis, C.~J., Eyles, C.~J., {et~al.} 2007, The Astrophysical
  Journal, 668, L79

\bibitem[{Wang {et~al.}(2024)Wang, Fatemi, Holmstr{\"o}m, Nilsson, Futaana, \&
  Barabash}]{wang2024mnras}
Wang, X.-D., Fatemi, S., Holmstr{\"o}m, M., {et~al.} 2024, Monthly Notices of
  the Royal Astronomical Society, 527, 12232

\bibitem[{{Williamson, H. N.} {et~al.}(2022){Williamson, H. N.}, {Nilsson, H.},
  {Stenberg Wieser, G.}, {Moeslinger, A.}, \& {Goetz, C.}}]{williamson2022aa}
{Williamson, H. N.}, {Nilsson, H.}, {Stenberg Wieser, G.}, {Moeslinger, A.}, \&
  {Goetz, C.} 2022, A\&A, 660, A103

\bibitem[{Yogesh {et~al.}(2023)Yogesh, Chakrabarty, \&
  Srivastava}]{yogesh_2023}
Yogesh, Chakrabarty, D., \& Srivastava, N. 2023, Monthly Notices of the Royal
  Astronomical Society: Letters, 526, L13

\bibitem[{Zhao {et~al.}(2019)Zhao, Li, Feng, Wu, Li, \& Zhao}]{zhao_2019}
Zhao, G.~Q., Li, H., Feng, H.~Q., {et~al.} 2019, The Astrophysical Journal,
  884, 60

\bibitem[{Ďurovcová {et~al.}(2017)Ďurovcová, Šafránková, Němeček, \&
  Richardson}]{durovcova_2017}
Ďurovcová, T., Šafránková, J., Němeček, Z., \& Richardson, J.~D. 2017,
  The Astrophysical Journal, 850, 164

\end{thebibliography}




\end{document}